\documentclass[aps,pra,twocolumn,groupedaddress,showpacs]{revtex4}
\usepackage{amsmath}
\usepackage{amssymb}
\usepackage{graphicx}

\begin{document}


\title{Two-dimensional Fourier-transform Spectroscopy of Potassium Vapor}


\author{X. Dai, A. D. Bristow, D. Karaiskaj}
\author{S. T. Cundiff}
\email{cundiffs@jila.colorado.edu}
\affiliation{JILA, University of Colorado and National Institute of Standards and Technology, Boulder, Colorado 80309-0440}

\date{\today}

\begin{abstract}
Optical two-dimensional Fourier-transformed (2DFT) spectroscopy is used to study the coherent optical response of potassium vapor in a thin transmission cell. Rephasing and non-rephasing spectra of the $D_{1}$ and $D_{2}$ transitions are obtained and compared to numerical simulations. Calculations using the optical Bloch equations gives very good agreement with the experimental peak strengths and line shapes. Non-radiative Raman-like coherences are isolated using a different 2DFT projection. Density-dependent measurements show distortion of 2DFT spectra due to pulse propagation effects.
\end{abstract}


\maketitle

\section{Introduction}
Vapors of alkali metal atoms are ideal for studying coherent light-matter interactions because the single outer electron results in a relatively simple optical spectrum with isolated lines. Alkali metals are the atom of choice in many areas of atomic physics such as ultracold atoms \cite{Bloch2008} and electromagnetically induced transparency \cite{Fleischhauer2005}, both of which are currently areas of extensive activity. While most studies have been carried out using continuous wave lasers and frequency domain methods, alkali atoms have also been studied using transient optical techniques. For example, tri-level \cite{Flusberg1978} and two-photon \cite{Flusberg1978} photon-echoes were first observed in a sodium vapor. Propagation effects \cite{KINROT1994, KINROT1995}, quantum interference \cite{GOLUB1986, KINROT1994, Shen2007} and non-Markovian dynamics \cite{Lorenz2005, Lorenz2008a} have been studied in a potassium vapor. Recently, rubidium vapor has been used as a model system for developing optical two-dimensional Fourier transform (2DFT) spectroscopy \cite{Tian2003,Vaughan2007,Tekavec2007}.

The technique of 2DFT spectroscopy was originally developed in nuclear magnetic resonance \cite{1987Ernst_Book}. Implementation in the optical region of the spectrum was first proposed using a Raman excitation scheme to study molecular vibrations \cite{1993Tanimura_JCP}. Currently, 2DFT techniques using infrared excitation to study molecular vibrations are widely used \cite{Cho2008,Fayer2009,Elsaesser2009}. 2DFT techniques using near-IR to visible excitation is also used to study electronic excitations in molecules \cite{Ginsberg2009} and semiconductor nanostructures \cite{Cundiff2009}. 2DFT spectroscopy explicitly measures the phase of a transient four-wave mixing (TFWM) signal during two time periods of a three-pulse excitation sequence \cite{Bristow2009a}. The resulting data is Fourier transformed onto a two-dimensional frequency plane, thus separating the electronic interactions between photoexcited states.

In this paper, we present a detailed study of the coherent optical response of potassium vapor in a thin transmission cell using optical 2DFT spectroscopy. Data is acquired for the standard rephasing and non-rephasing configurations, as well as for the projection that isolates the non-radiative ``Raman'' coherence \cite{Yang2008c}. We observe peak shapes and amplitudes associated with two coupled $D$ lines. By increasing the number density, we can observe effects due to propagation. The rephasing and non-rephasing spectra are compared to numerical simulations based on the optical Bloch equations.

\section{Experimental}

\begin{figure}
\includegraphics[width=3.0in]{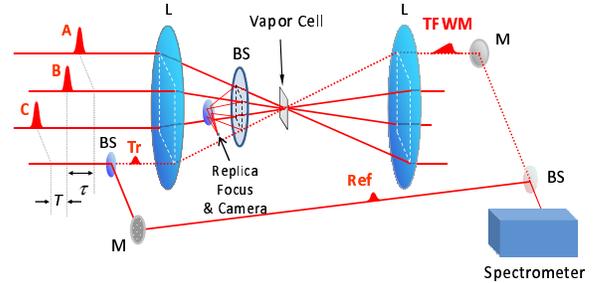}
\caption{\label{fig:Scheme}(color online) Schematic diagram of the experimental setup. Notation: L:lens, BS: beam splitter, M:mirror.}
\end{figure}

A schematic of the experiment is given in Fig.~\ref{fig:Scheme}. A mode-locked Ti:sapphire laser supplies ${\sim}200fs$ pulses, centered at 768.2 nm, to the multidimensional optical nonlinear spectrometer \cite{Bristow2009a}. The spectrometer splits the Ti:sapphire pulses  into four phase-locked replicas that are aligned in the box geometry. The pulses are focused on the same spot of a home-made alkali vapor cell. Three pulses (A, B and C) generate a TFWM signal. Time delays between the pulses are denoted $\tau$, $T$ and $t$ between the first and second, the second and third and the third pulse and signal respectively. Pulse four provides the phase-stabilized tracer (Tr) and reference (Ref) separately: the tracer co-propagates with the TFWM signal and is useful for alignment, but is blocked during the 2DFT measurements; the reference beam is routed around the vapor cell and recombined with the signal, producing an interferogram that is recorded by a cooled CCD spectrometer. The phase of the signal field is determined by an all-optical method, which is realized by careful measurement of relative phases between all laser pulses and the TFWM signal \cite{Bristow2008}. Further experimental details can be found elsewhere \cite{Bristow2009a, Bristow2009}.

The TFWM signal, $E_s \sim E_A^*E_B E_C$, is generated in the phase-matched direction $\mathbf{k}_{s} = -\mathbf{k}_{A}+\mathbf{k}_{B}+\mathbf{k}_{C}$ where $E_A^*$ is the conjugated pulse. The subscripts, $A, B$ or $C$ only denote direction, not time ordering.  If $E_A$ arrives first the resulting 2DFT spectrum, $S_{I}(\tau,T,t)$, is known as ``rephasing'' because the dephasing due to inhomogeneous broadening is canceled. In a TFWM experiment, this time ordering produces a photon echo. If $E_A$ arrives second, the cancelation of dephasing due to inhomogeneous broadening does not occur and the resulting $S_{II}(\tau,T,t)$ spectra are referred to as ``nonrephasing''. $S_{III}(\tau,T,t)$ spectra, where $E_A$ arrives last, are sensitive to two-quantum resonances \cite{Stone2009, Yang2008}. The time resolved spectra $S_{i}(\tau,T,t)$ are usually Fourier transformed with respect to two time delays, while the third delay is fixed. The time delays $\tau$ and $T$ are defined based on time ordering, not direction, and thus are strictly positive.

Reflection cells designed for high temperature vapor experiments \cite{Lorenz2008} can not be used in our multidimensional spectroscopy apparatus because the signal is detected in transmission \cite{Bristow2009a}. Furthermore, in a reflection geometry the phase of the signal depends on the position of the interface producing the reflection. Therefore, a thin transmission cell for alkali atomic vapor has been designed and manufactured for the current experiment. The cell body is machined from high-grade titanium. Two sapphire windows are separately diffusion bonded to the titanium body. The gap between two windows can be adjusted before the transmission cell is loaded with solid potassium \cite{Lorenz2008}. The vapor temperature is controlled by a heater attached to the transmission cell, up to a maximum of $800^{\circ}\mathrm{C}$. However, for the measurement reported here, the cell temperature is varied from $210^{\circ}\mathrm{C}$ to $270^{\circ}\mathrm{C}$, ensuring low absorption and number densities in the range $2.03\times 10^{20} m^{-3}$ to $1.87\times 10^{21} m^{-3}$ \cite{Fiock1926}. During the measurements, the absorption spectrum of the potassium vapor at each temperature is recorded to estimate the gap size between two windows by comparing experimental spectra with calculated absorption spectra.

The measured linear absorption and the energy level diagram for atomic potassium are shown in Fig.~\ref{fig:K2DExpl}(a) and (b) respectively. The laser excites both $D_{1}$ and $D_{2}$ transitions and there are no resonant two-photon transitions. Fig.~\ref{fig:K2DExpl}(a) also shows the calculated linear absorption spectrum at $250^{\circ}\mathrm{C}$. At this temperature the resonance broadened linewidths of the $D_{1}$ and $D_{2}$ transitions are below the spectrometer resolution. A ${\sim}1500$ Torr argon buffer gas is added to the cell to induce collision broadening. The argon  reduces the absorbance ${\alpha}L$ below 1 into the weak absorption regime. In the calculation, the spectral linewidth is the sum of resonance from potassium-potassium interactions \cite{LEWIS1971} and collisional broadening with argon \cite{LWIN1978}. The gap size, $L$, is estimated to be $19\mu$m at $250^{\circ}\mathrm{C}$ and varies slightly with temperature over the range explored.

\section{Results and discussion}

\begin{figure}
\includegraphics[width=3.0in]{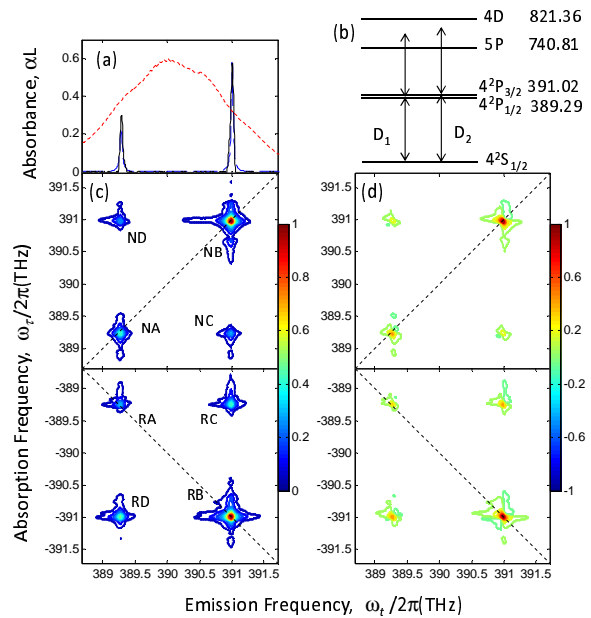}
\caption{\label{fig:K2DExpl}(color) (a) Excitation pulse spectrum (dotted) and experimental (solid) and calculated (dashed) linear absorption of potassium vapor at $250^{\circ}\mathrm{C}$. $\alpha$ is the absorption coefficient and $L$ is the gap between the two sapphire windows. (b) The relevant energy levels of atomic potassium, with frequencies in THz. (c) Experimental amplitude $S_{I}(\omega_{\tau}, T, \omega_{t})$ (bottom) and $S_{II}(\omega_{\tau}, T, \omega_{t})$ (top) spectra of potassium vapor. (d) Experimental real-part $S_{I}(\omega_{\tau}, T, \omega_{t})$ (bottom) and $S_{II}(\omega_{\tau}, T, \omega_{t})$ (top) spectra.The spectra are normalized to the most intense peak.}
\end{figure}

Figure~\ref{fig:K2DExpl}(c) and (d) show the amplitude and real-part of $S_{I}(\omega_{\tau}, T, \omega_{t})$ and $S_{II}(\omega_{\tau}, T, \omega_{t})$ 2DFT spectra of potassium vapor at $250^{\circ}\mathrm{C}$. The emission frequency, $\omega_t$ is defined to be positive, thus for $S_{I}(\omega_{\tau}, T, \omega_{t})$, the signal appears at negative $\omega_{\tau}$ because the first pulse is conjugated.  The $S_{I}(\omega_{\tau}, T, \omega_{t})$ and $S_{II}(\omega_{\tau}, T, \omega_{t})$ spectra can be used to determine homogeneous and inhomogeneous linewidths \cite{Faeder1999, Kuznetsova2007}. The horizontal axis of the spectrum is generated directly from the recorded interferogram between the reference pulse and the TFWM signal \cite{Lepetit1995} at each delay step between the pulses A and B. The vertical axis of the spectrum is generated by fast Fourier transform with respect to $\tau$. During the measurements, $T$ was set to $580fs$, which is equal to $1/(f_{D2}-f_{D1})$ where $f_{D1}$ and $f_{D2}$ are the transitions frequencies corresponding to the transitions $D_{1}$ and $D_{2}$ respectively.

\begin{figure}
\includegraphics[width=3.0in]{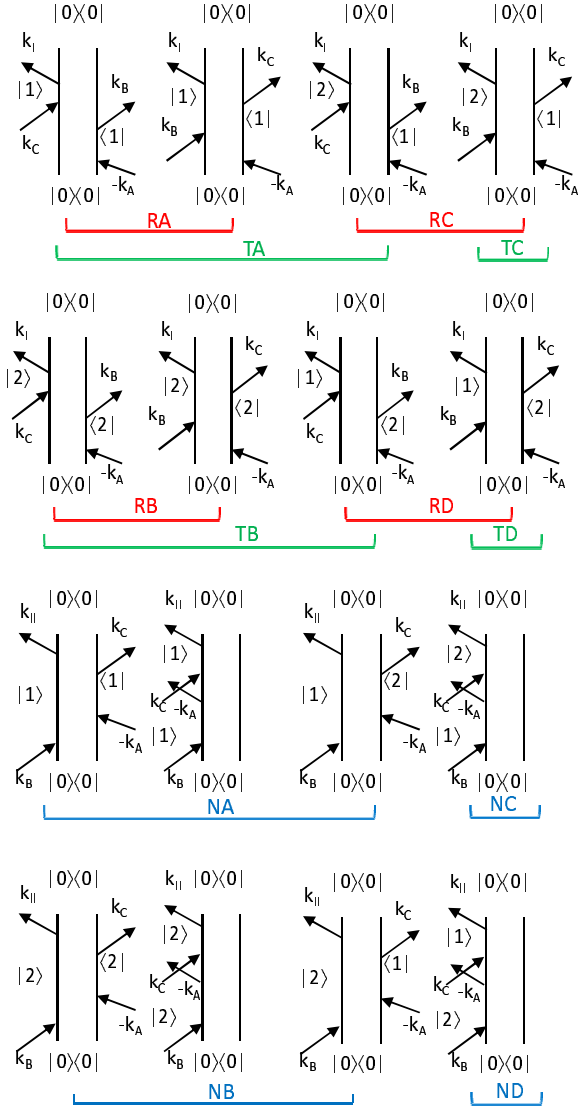}
\caption{\label{fig:FeynmanDiag} (color online) Double-sided Feynman diagrams for all possible Liouville-space pathways contributing to the third-order optical response of a V-type system. The pathways for the $S_{I}(\omega_{\tau}, T, \omega_{t})$ and $S_{I}(\tau, \omega_{T}, \omega_{t})$ measurements are listed in top two rows; while the pathways for the $S_{II}(\omega_{\tau}, T, \omega_{t})$ measurement are listed in bottom two rows. The state numbers, 0, 1 and 2, correspond to $4^2S_{1/2}$, $4^2P_{1/2}$ and $4^2P_{3/2}$ of potassium atoms.}
\end{figure}

Each spectrum has two diagonal peaks, which correspond to the $D_{1}$ and $D_{2}$ transitions, and two off-diagonal peaks, which represent coupling between the two resonances. However, the relative peak strengths are different in $S_{I}(\omega_{\tau}, T, \omega_{t})$ and $S_{II}(\omega_{\tau}, T, \omega_{t})$ spectra. This difference can be qualitatively explained by the double-sided Feynman diagrams shown in Fig.~\ref{fig:FeynmanDiag}. Double-sided Feynman diagrams are useful for illustrating all possible Liouville-space pathways in the systems \cite{Mukamel2000}. Two pathways contribute to each peak in the $S_{I}(\omega_{\tau}, T, \omega_{t})$ spectra; whereas for $S_{II}(\omega_{\tau}, T, \omega_{t})$ three pathways contribute to each diagonal peak (NA or NB) and only one pathway to each cross peak (NC or ND). Generally, the peak strengths are proportional to the sum over participating diagrams while the strength of individual diagram is proportional to ${\prod}\mu_{i}$, where index $i$ runs from 1 to 4 and $\mu_{i}$ is the transition dipole moment for each excitation step in each double-sided Feynman diagram. The ratio between the transition dipole moment for $D_{2}$ transition, $\mu_{D2}$, to the dipole moment for $D_{1}$ transition, $\mu_{D1}$, is approximately $\sqrt{2}:1$. Assuming each quantum pathway has the same contribution to the peak strength, the peak strength ratio RA:RC:RD:RB is proportional to $2 \mu_{D1}^4:2 \mu_{D1}^2 \mu_{D2}^2:2 \mu_{D2}^2 \mu_{D1}^2:2 \mu_{D2}^4$, or relative values of 1:2:2:4. Similarly, the peak strength ratio in the $S_{II}(\omega_{\tau}, T, \omega_{t})$ measurement NA:NC:ND:NB is equal to 2:1:1:5, which agrees well with the experimental result.

\begin{figure}
\includegraphics[width=3.0in]{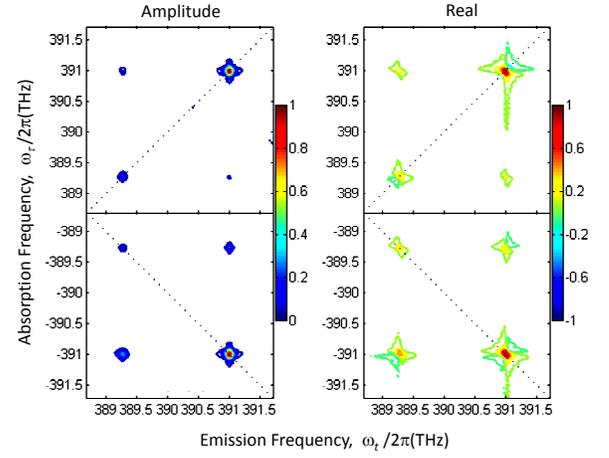}
\caption{\label{fig:K2DThry}(color) Calculated amplitude (left) and real-part (right) 2DFT spectra of potassium vapor for $S_{I}(\omega_{\tau}, T, \omega_{t})$ (bottom) and $S_{II}(\omega_{\tau}, T, \omega_{t})$ (top) measurements. The spectra are normalized to the most intense peaks.}
\end{figure}

The interaction between light and a closed two-level system can be described by the optical Bloch equations \cite{Li2006}:
\begin{align}
	\dot{\rho}_{00}& = -\dot{\rho}_{11}\\
	\dot{\rho}_{11}& = -{\gamma}^{sp}\rho_{11}+\frac{i}{\hbar}\mu_{01}E(\rho_{01}-\rho_{10})\\
	\dot{\rho}_{01}& = -{\gamma}^{ph}\rho_{01}+i\omega_{01}\rho_{01}+\frac{i}{\hbar}\mu_{01}E(\rho_{11}-\rho_{00})
\end{align}
where $\rho_{00}$ and $\rho_{11}$ are density matrix elements for the ground and excited state population respectively, and $\rho_{01}$ is the off-diagonal term of the density matrix. $\gamma^{sp}$ and $\gamma^{ph}$ are the population relaxation and dephasing rates respectively, $\mu_{01}$ is the transition dipole moment, $\omega_{01}$ is the resonant transition frequency and $E$ is the electric field amplitude. We extended the equations to a multi-level system and numerically solved them \cite{Shacklette2002, Li2006}. In the calculation, $\gamma^{ph}$ is the total linewidth due to resonance and collision broadening. The induced polarization of the system is $P=NTr(\mu \rho)$. Then the electric field of the emitted signal is determined from the third order polarization \cite{Li2006}:
\begin{equation}
E(\tau,T,\omega_{t})=\frac{2\pi L}{n(\omega_{t})c}i\omega_{t}P^{(3)}(\tau,T,\omega_{t})
\end{equation}
where $L$ is the sample thickness, $n(\omega_{t})$ is the index of refraction of the sample, and $c$ is the speed of light in vacuum. $S_{I}(\omega_{\tau}, T, \omega_{t})$ and $S_{II}(\omega_{\tau}, T, \omega_{t})$ 2DFT spectra are defined as the Fourier transform with respect to delay $\tau$:
\begin{equation}
S_{I,II}(\omega_{\tau},T,\omega_{t}) = \int_{-\infty}^{\infty} E(\tau,T,\omega_{t})e^{i \omega_{\tau} \tau}\mathrm{d}\tau
\end{equation}

The numerical simulations of the $S_{I}(\omega_{\tau}, T, \omega_{t})$ and $S_{II}(\omega_{\tau}, T, \omega_{t})$ spectra shown in Fig.~\ref{fig:K2DThry} are in very good agreement with the experiment; correct peak strengths and symmetric star-like line shapes are reproduced in the amplitude spectra, indicating the dominance of homogeneous broadening \cite{Faeder1999}. As expected the lineshapes are absorptive, which in contrast to semiconductors where many-body interactions result in dispersive lineshapes \cite{Cundiff2009,Bristow2009,Li2006}.

\begin{figure}
\includegraphics[width=3.0in]{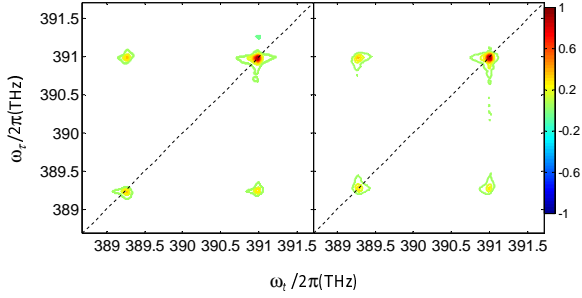}
\caption{\label{fig:Absorptive}(color) Experimental (left) and theoretical (right) purely absorptive spectra obtained by the addition of the $S_{I}(\omega_{\tau}, T, \omega_{t})$ and $S_{II}(\omega_{\tau}, T, \omega_{t})$ spectra.}
\end{figure}

There are both absorptive and dispersive contributions in the $S_{I}(\omega_{\tau}, T, \omega_{t})$ and $S_{II}(\omega_{\tau}, T, \omega_{t})$ spectra. Purely absorptive contributions can be isolated by adding the $S_{I}(\omega_{\tau}, T, \omega_{t})$ and $S_{II}(\omega_{\tau}, T, \omega_{t})$ signals together\cite{Khalil2003, Tekavec2007}. The experimental and theoretical purely absorptive spectra are shown in Fig.~\ref{fig:Absorptive}. Compared with the $S_{I}(\omega_{\tau}, T, \omega_{t})$ and $S_{II}(\omega_{\tau}, T, \omega_{t})$ spectra, the disappearance of dispersive wings in the purely absorptive spectra results in a higher resolution of spectral features.

\begin{figure}
\includegraphics[width=2.0in]{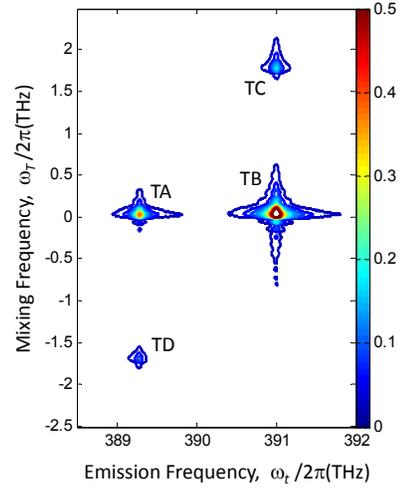}
\caption{\label{fig:K2DTscan}(color) Experimental amplitude $S_{I}(\tau,\omega_{T},\omega_{t})$ spectrum.}
\end{figure}

Cross peaks in $S_{I}(\omega_{\tau}, T, \omega_{t})$ have contributions from ground-state bleaching and excited-state emission pathways. These two pathways differ during the delay time $T$. After the second excitation pulse, the ground-state bleaching pathway leaves the system in the ground state, while the excited-state emission or Raman-like pathway (labeled TC or TD in Fig.~\ref{fig:FeynmanDiag}) leaves the system in a coherent superposition of $4^2P_{1/2}$ and $4^2P_{3/2}$. $S_{I}(\tau,\omega_{T},\omega_{t})$ spectra can be used to isolate Raman coherences terms, projecting the 2DFT signal along $\omega_{T}$ and $\omega_{t}$ instead of $\omega_{\tau}$ and $\omega_{t}$ \cite{Yang2008c}. Technically, the measurement is achieved by scanning the third pulse C while keeping the time delay $\tau$ between first two pulses constant. The spectrum is then obtained by fast Fourier transform of time domain signal with respect to $T$. Figure \ref{fig:K2DTscan} shows an experimental spectrum of $S_{I}(\tau,\omega_{T},\omega_{t})$, with $\tau = 0$ in the measurement. In the spectrum, two strong peaks are observed along the zero mixing energy and two side peaks show up at the energy position of TC $(E_{D2}, E_{D2}-E_{D1})$ and TD $(E_{D1}, E_{D1}-E_{D2})$, where $E_{D1}$ and $E_{D2}$ are the photon energies corresponding to the transitions $D_{1}$ and $D_{2}$ separately. The amplitude of the spectrum is normalized according to the strength of the strongest peak, however, the spectrum is displayed with a color scale of 0 to 0.5 to emphasize two side peaks. It is clear that the two quantum pathways for Raman coherences (TC and TD) can be isolated as two side peaks in the $S_{I}(\tau,\omega_{T},\omega_{t})$ spectrum. Weak asymmetry between TC and TD features is observed, where the stronger peak has the same emission energy as the stronger zero mixing energy feature. This asymmetry is also observed in semiconductors \cite{Yang2008c}.

The dephasing rate of the Raman coherence is related to the dephasing rates of the optical transitions by \cite{Spivey2008}
\begin{equation}
\gamma_{D2-D1} = \gamma_{D2} + \gamma_{D1} - 2 R_{ph} (\gamma_{D1} \gamma_{D2})^{1/2}
\end{equation}
where $\gamma_{D2-D1}$, $\gamma_{D2}$ and $\gamma_{D1}$ are the dephasing rates for the Raman coherences, the $D_{2}$ and $D_{1}$ transitions respectively. The coefficient $R_{ph}$ describes the degree of correlation between fluctuations of the two transitions. The dephasing rates of $D_{2}$ and $D_{1}$ extracted from the $S_{I}(\omega_{\tau}, T, \omega_{t})$ measurement are 0.128 THz and 0.118 THz. The dephasing rates for the Raman coherences at the energy position of TC and TD are 0.127 THz and 0.140 THz respectively so the correlation coefficients $R_{ph}$ are 0.45 and 0.43 respectively. The positive correlation coefficients means that the energies of the $D_{2}$ and $D_{1}$ lines simultaneously shift in the same direction relative to each other during the collisional scattering, as expected.

\begin{figure}
\includegraphics[width=3.0in]{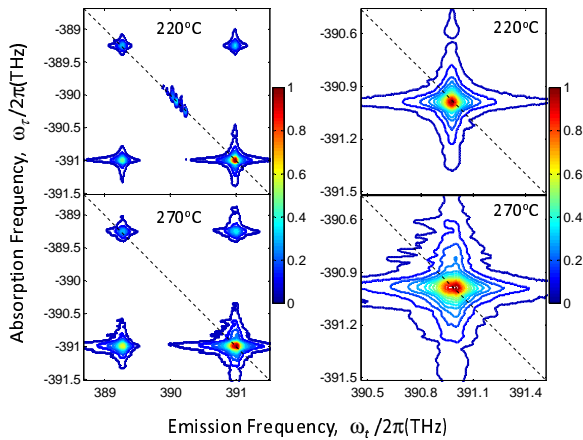}
\caption{\label{fig:K2DPropagation}(color) Distortion of the peak shapes due to propagation effects. Left two panels are the $S_{I}(\omega_{\tau}, T, \omega_{t})$ spectra of the potassium vapor at $220^{\circ}\mathrm{C}$ (top) and $270^{\circ}\mathrm{C}$ (bottom). Right two panels are the zoomed pictures of the diagonal peaks corresponding to the $D_{2}$ transition at $220^{\circ}\mathrm{C}$ (top) and $270^{\circ}\mathrm{C}$ (bottom).}
\end{figure}

As in other types of spectroscopy, the lineshapes in 2DFT spectroscopy can be affected by the propagation of the excitation laser pulses and signals in the resonant medium. Propagation effects in TFWM experiments of alkali vapors have been discussed \cite{KINROT1995}, as have distortions of 2DFT spectra due to propagation effects \cite{Keusters2003, Keusters2004, Yetzbacher2007, Kwak2008}. When the number density of atomic potassium is increased by raising the temperature, the absorption also strengthens. Distortion of 2DFT spectra peaks due to propagation can then be observed. $S_{I}(\omega_{\tau}, T, \omega_{t})$ spectra of potassium vapor at $220^{\circ}\mathrm{C}$ and $270^{\circ}\mathrm{C}$ are shown in Fig.~\ref{fig:K2DPropagation}, where the absorbance $ \alpha L$ is ${\sim}0.2$ and ${\sim}1.3$ for the $D_{2}$ transition, respectively. At the lower temperature, the diagonal $D_{2}$ feature is a symmetric star shape, while at high temperature it is elongated along the emission photon axis. This change has been predicted \cite{Keusters2003, Keusters2004, Yetzbacher2007}, although the expected change in peak strength ratio is not observed \cite{Keusters2003}.

\section{Conclusion}

$S_{I}(\omega_{\tau}, T, \omega_{t})$ and $S_{II}(\omega_{\tau}, T, \omega_{t})$ 2DFT spectra for potassium vapor have been obtained using a multidimensional nonlinear spectrometer. Numerical simulation based on the optical Bloch equations reproduce the experimental spectra giving excellent agreement. Purely absorptive spectra are obtained by summing $S_{I}(\omega_{\tau}, T, \omega_{t})$ and $S_{II}(\omega_{\tau}, T, \omega_{t})$. We also experimentally isolated Raman coherences by projecting the 2DFT signal along $\omega_{T}$ and $\omega_{t}$ instead of $\omega_{\tau}$ and $\omega_{t}$. Finally, the expected distortion of 2DFT spectra due to propagation effect is observed at higher vapor temperatures.

2DFT spectroscopy has been proved to be a powerful tool to show interactions and coupling in complex systems, it can also be used to investigate simple atomic and small molecular systems. As we have demonstrated, 2DFT spectroscopy of simple systems can be useful to validate the various techniques and theory for 2DFT spectroscopy.

\begin{acknowledgments}
The authors thank T. Asnicar and H. Green for technical assistance. The financial support is provided by NIST and the National Science Foundation's Physics Frontier Center Program. S.T.C. is a staff member in the NIST Quantum Physics Division.
\end{acknowledgments}

\end{document}